\documentclass[final,3p,onecolumn,times]{elsarticle}

\usepackage{amssymb}
\usepackage{amsmath}
\usepackage{amsthm}
\usepackage[utf8]{inputenc}
\usepackage{physics}
\usepackage{bm}
\usepackage{microtype}
\usepackage{hyperref}
\usepackage{color}

\journal{Physica A: Statistical Mechanics and its Applications}

\newcommand{\mean}[1]{\left\langle #1 \right\rangle}
\newcommand{\st}{\text{s}}
\newcommand{\hcs}{\text{HCS}}
\newcommand{\ini}{\text{i}}

\newcommand{\calR}{\mathcal{R}}
\newcommand{\cool}{\text{cool}}

\biboptions{sort&compress}

\begin{document}

\begin{frontmatter}



\title{Optimal preparation and reachable-state constraints in the Mpemba effect}


\author[label1]{C.~Ríos-Monje} 
\author[label2]{A.~Patrón-Castro}
\author[label1]{C.~A.~Plata}
\author[label3]{B.~Sánchez-Rey}
\author[label1]{A.~Prados}
\affiliation[label1]{organization={Física Teórica, Multidisciplinary Unit for Energy Science, Universidad de Sevilla},
            addressline={Apartado de Correos 1065}, 
            city={Sevilla},
            postcode={E-41080}, 
            country={Spain}}
\affiliation[label2]{organization={Department of Physics, Simon Fraser University}, 
            city={Burnaby},
            postcode={V5A 1S6}, 
            state={British Columbia},
            country={Canada}}
\affiliation[label3]{organization={Departamento de Física Aplicada I, E.P.S., Multidisciplinary Unit for Energy Science, Universidad de Sevilla}, 
            addressline={Virgen de África 7}, 
            city={Sevilla},
            postcode={E-41011}, 
            country={Spain}}


\begin{abstract}


The Mpemba effect, whereby an initially hotter system relaxes faster than a colder one towards a common final state, is often analysed within the kinetic framework by assuming non-stationary initial conditions that are selected a priori. Here, we revisit this viewpoint by explicitly incorporating the aging protocol used to prepare those states. Focusing on uniformly heated granular fluids, we formulate the preparation stage as an optimal-control problem in which the energy injection is tuned to generate the initial conditions that maximise or minimise the subsequent relaxation rate. Within the first Sonine approximation, this optimisation reduces to extremising the excess kurtosis of the velocity distribution function, the quantity controlling the cooling rate at fixed temperature. Applying Pontryagin's maximum principle, we show that the optimal preparation protocol is always a one-bang protocol and determine the corresponding extremal values of the excess kurtosis. Most importantly, we find that the stochastic thermostat imposes non-trivial reachable-state constraints: the accessible non-Gaussianities are bounded by those of the homogeneous cooling state, thereby limiting the relaxation-rate asymmetry and the maximum attainable Mpemba effect. These results demonstrate that the strength of the kinetic Mpemba effect cannot be disentangled from the accessibility of the underlying non-equilibrium states. More generally, our work establishes a connection between anomalous relaxation, optimal control, and state accessibility in non-equilibrium systems.
\end{abstract}




\end{frontmatter}



\section{Introduction}
\label{sec:intro}

The Mpemba effect, i.e. the counterintuitive phenomenon whereby a system initially prepared at a higher temperature may relax faster than an identical system at a lower temperature~\cite{mpemba_cool_1969}, has attracted renewed attention in recent years as a paradigmatic example of anomalous relaxation in non-equilibrium systems~\cite{patron_non-equilibrium_2023,teza_speedups_2026}. Although originally reported in the context of water freezing, it is now understood as a generic feature of relaxation processes in many different systems~\cite{lu_nonequilibrium_2017,lasanta_when_2017,klich_mpemba_2019,kumar_exponentially_2020,kumar_anomalous_2022,tian_experimental_2025,song_mpembalike_2026}. In this broader perspective, the Mpemba effect—and its inverse—is intimately connected with the existence of fast relaxation routes to equilibrium, raising fundamental questions about the dynamical organisation of non-equilibrium states and their approach to stationarity~\cite{lapolla_faster_2020,gal_precooling_2020,prados_optimizing_2021,patron_thermal_2022,teza_eigenvalue_2023,teza_relaxation_2023,ibanez_heating_2024,van_vu_thermomajorization_2025,summer_resource-theoretical_2026}. Understanding such anomalous relaxation pathways ultimately requires not only characterising non-equilibrium states, but also determining which of them can be reached through physically admissible preparation protocols.

Two main theoretical frameworks have been put forward to analyse the Mpemba effect. On the one hand, the entropic—or stochastic—approach defines the Mpemba effect in terms of the temporal evolution of a suitable distance in probability space from an initial equilibrium state to the targeted final equilibrium state, leading to a description based on entropic or information-theoretic measures~\cite{lu_nonequilibrium_2017,baity-jesi_mpemba_2019,klich_mpemba_2019,busiello_inducing_2021,kumar_exponentially_2020,chetrite_metastable_2021,carollo_exponentially_2021,kumar_anomalous_2022,schwarzendahl_anomalous_2022,biswas_mpemba_2023,tian_experimental_2025,van_vu_thermomajorization_2025}. On the other hand, the kinetic—or thermal—approach focuses on the crossing of macroscopic observables such as the kinetic temperature of fluids and emphasises the dynamical role of non-steady initial conditions~\cite{lasanta_when_2017,torrente_large_2019,santos_mpemba_2020,biswas_mpemba_2020,mompo_memory_2021,takada_mpemba_2021,gomez_gonzalez_mpemba-like_2021,patron_strong_2021,megias_mpemba-like_2022}. Within this latter framework, the Mpemba effect emerges from a suitable preparation of the initial non-equilibrium state, such that the relaxation rate of the initially hotter sample is enhanced with respect to that of the colder one. Consequently, the magnitude of the effect is closely tied to the set of non-equilibrium states that can be generated during the preparation stage. It must be noted that the two approaches are not equivalent in general, and situations exist where one formulation predicts a Mpemba effect while the other does not~\cite{megias_thermal_2022}.

From the kinetic viewpoint, the Mpemba effect constitutes a genuine memory effect, since the relaxation dynamics depend not only on the instantaneous macroscopic variables but also on the previous history of the system, i.e., on the aging protocol that led to the initial state. This feature places it in close conceptual relation with other non-equilibrium memory phenomena, most notably the Kovacs effect, for which the dependence on the thermal history is well established~\cite{kovacs_transition_1963,kovacs_isobaric_1979,prados_kovacs_2010,bouchbinder_nonequilibrium_2010}. In granular and glassy systems, such memory effects arise from the presence of additional internal variables, which encode the deviation of the instantaneous state of the system from equilibrium and thereby influence the subsequent evolution~\cite{bertin_kovacs_2003,prados_kovacs-like_2014,trizac_memory_2014,ruiz-garcia_kovacs_2014,lulli_kovacs_2019,militaru_kovacs_2021}. From this perspective, the preparation stage can be viewed as a means of steering these internal variables, and hence of controlling the ensuing relaxation dynamics.

Some recent work has investigated the role of an aging window in connection with time-delayed versions of Newton's law of cooling~\cite{santos_mpemba_2024,santos_time-delayed_2025,santos_mpemba_2026}. Nevertheless, in the kinetic formulation of the Mpemba effect, the role of the aging protocol itself has received comparatively little attention. In this respect, it is worth contrasting the standard single-step Mpemba setups (as considered in both entropic and kinetic approaches) with the recently introduced Pontus--Mpemba protocols, in which the total time needed to reach the final target state—including the preparation time required to reach an intermediate non-stationary state—is explicitly taken into account~\cite{nava_pontus-mpemba_2025,peluso_optimal_2026}. 
By construction, Pontus--Mpemba protocols shift the emphasis from selecting favourable initial conditions \emph{a posteriori} to assessing what can actually be achieved under physically admissible preparation dynamics, thereby making state accessibility and preparation cost central ingredients~\cite{nava_pontus-mpemba_2025}.  This naturally raises the question of which non-equilibrium states can be reached through a given preparation mechanism and, consequently, what is the largest Mpemba effect that can be realised under the corresponding dynamical constraints. 

Motivated by this protocol-based viewpoint, the aim of this work is to determine the optimal aging protocol that maximises the Mpemba effect within the kinetic framework. More specifically, we seek preparation protocols that maximise (minimise) the relaxation rate of the initially hotter (colder) sample and establish the corresponding limits imposed by the accessible set of non-equilibrium states. This optimisation problem is naturally formulated within optimal control theory~\cite{pontryagin_mathematical_1987,liberzon_calculus_2012}, which provides systematic tools to identify extremal dynamical trajectories in driven systems~\cite{guery-odelin_driving_2023}. For concreteness, we focus on uniformly heated granular fluids driven by a stochastic thermostat~\cite{van_noije_velocity_1998,van_noije_randomly_1999,montanero_computer_2000,garcia_de_soria_universal_2012}, the first system in which the Mpemba effect was theoretically predicted within the kinetic framework~\cite{lasanta_when_2017}. Interestingly, such stochastic driving emerges when a granular fluid is immersed in an elastic background fluid of much lighter particles but comparable density~\cite{santos_granular_2003,gomez_gonzalez_kinetic_2022}. In this way, the kinetic Mpemba effect is recast as a problem of optimal state preparation under dynamical reachability constraints.

A crucial ingredient in our analysis is the role played by non-Gaussianities in the velocity distribution function (VDF). In the kinetic description of granular gases, deviations from the Maxwellian distribution are systematically characterised through Sonine polynomial expansions, in which the leading correction is quantified by the excess kurtosis~\cite{van_noije_velocity_1998,montanero_computer_2000}. Within the first Sonine approximation, the relaxation rate at fixed temperature is directly governed by this quantity, increasing with the magnitude of the non-Gaussian correction. This observation allows us to recast the optimisation problem in terms of the maximisation or minimisation of the excess kurtosis through suitable aging protocols, thereby providing a simple physical interpretation of the control strategy. In particular, the extrema of the excess kurtosis define the boundaries of the non-equilibrium states that can be reached under a given driving mechanism and therefore determine the largest relaxation-rate asymmetry that can be achieved.

The remainder of the paper is organised as follows. In Section~\ref{sec:model}, we introduce the kinetic description of the system and put forward the evolution equations for the kinetic temperature and the excess kurtosis. Section~\ref{sec:optimal-control} formulates the aging stage as an optimal-control problem and derives the optimal preparation protocols together with the extremal accessible values of the excess kurtosis. These theoretical predictions are checked against numerical simulations of the granular-gas dynamics in Section~\ref{sec:numerics}. Finally, Section~\ref{sec:discussion} discusses the implications of our results for state accessibility, memory effects, and anomalous relaxation, and outlines several directions for future work.


\section{The Model}
\label{sec:model}

Our model is a uniformly heated granular gas of $d$-dimensional hard
spheres with mass $m$ and diameter $\sigma$. The gas particles undergo
inelastic collisions, with restitution coefficient
$\alpha$, and additionally are subjected to a stochastic thermostat, i.e. a white noise force with correlation amplitude $m^{2}\xi^{2}$. The system remains spatially homogeneous if initially so,
and its mesoscopic state is then described by the dynamical
evolution of its VDF. If the number density $n$ is low enough, the Boltzmann-Fokker-Planck equation for the VDF accurately describes the system dynamics~\cite{van_noije_velocity_1998,van_noije_randomly_1999,montanero_computer_2000}. 

Here, we carry our theoretical analysis in the first Sonine approximation for the Boltzmann-Fokker-Planck equation. In this description, the state of the gas is characterised by the granular (kinetic) temperature $T$, which is proportional to the average energy per particle, and the excess kurtosis $a_2$, defined as
\begin{equation}
  \label{eq:T-a2-def}
  T\equiv \frac{m \mean{v^{2}}}{d}, \qquad a_{2}\equiv\frac{d}{d+2}\frac{\mean{v^{4}}}{\mean{v^{2}}^{2}}-1,
\end{equation}
which measures the non-Gaussianities in the VDF. It is worth stressing that the non-Gaussianities, which play the role of internal variables controlling relaxation and memory effects in granular gases, are well captured by the excess kurtosis in many physical situations~\cite{van_noije_velocity_1998,montanero_computer_2000,garcia_de_soria_universal_2012,trizac_memory_2014,prados_kovacs-like_2014,lasanta_when_2017,patron_kinetic_2024}.

For time-independent driving, the granular gas reaches a non-equilibrium steady state (NESS) in the long time limit: the energy loss in collisions and the energy input from the thermostat balance each other on average. The stationary values of $T$ and $a_2$ satisfy~\cite{van_noije_velocity_1998,montanero_computer_2000}
\begin{align}\label{eq:Ts-and-a2s} \chi\equiv T_\st^{3/2}&=\frac{m\xi^2}
{\zeta_0(1+\frac{3}{16}a_2^\st)}, &  \zeta_0&=\frac{2 n
\sigma^{d-1} \left(1-\alpha^2\right) \pi^{\frac{d-1}{2}}}{\sqrt{m}
d\,\Gamma(d/2)}, &
 a_{2}^{\st}&=\frac{16(1-\alpha)(1-2\alpha^2)}
{73+56d-24d\alpha-105\alpha+30(1-\alpha)\alpha^2}.
\end{align}

In the first Sonine approximation, $T$ and $a_{2}$ obey a closed set of equations, linear in $a_2$: both higher-order cumulants  and nonlinearities in $a_2$ are neglected, 
\begin{subequations}\label{eq:evol-with-dim}
\begin{align}
  \dot{T}&=\zeta_{0}\left[\chi\left(1+\frac{3}{16}a_{2}^{\st}\right)-T^{3/2}
  \left(1+\frac{3}{16}a_{2}\right)\right], \\
  \dot{a_{2}}&=\frac{2\zeta_{0}}{T}\left[\left(T^{3/2}-\chi\right)a_{2}+B\,T^{3/2}
  \left(a_{2}^{\st}-a_{2}\right)\right].
\end{align}
\end{subequations}
The parameter $B$ is only a function of $\alpha$ and $d$, namely
\begin{align}\label{eq:B-and-a2HCS}
  B=&\frac{a_{2}^{\hcs}}{a_{2}^{\hcs}-a_{2}^{\st}}, &
  a_{2}^{\hcs}&=\frac{16(1-\alpha)(1-2\alpha^{2})} {25
+2\alpha^{2}(\alpha-1)+ 24d+\alpha(8d-57)},
\end{align}
where $a_{2}^{\hcs}$ is the value of the excess kurtosis in the long-time time-dependent state reached by the system in the absence of a thermostat, $\chi=0$ (free cooling)---what is called the homogeneous cooling state (HCS). Note that both $a_{2}^{\st}$ and $a_{2}^{\hcs}$ depend on the value of the restitution coefficient $\alpha$. This dependence will be central to the optimal-control problem discussed below, since it determines the sign of the excess kurtosis and, ultimately, the structure of the accessible set of non-equilibrium states. We differentiate two regimes: $\alpha<\alpha_c\equiv 1/\sqrt{2}$, where both $a_{2}^{\st}$ and $a_{2}^{\hcs}$ are positive, and  $\alpha>\alpha_c$, where they are both negative.

In the following, we employ dimensionless variables: 
\begin{equation}
  \label{eq:nondim}
  T^{*}=T/T_{\ini}, \quad \chi^{*}=\chi/T_{\ini}^{3/2}, \quad \bm{v}^{*}=\sqrt{m/T_{\ini}}\,\bm{v}, \quad
  t^{*}=\zeta_{0}T_{\ini}^{1/2}t,
\end{equation}
where $T_{\ini}$ is the initial value of the
temperature, $T_{\ini}=T(t=0)$. Note that the excess
kurtosis is already dimensionless. In what follows, the asterisks are omitted to prevent cluttering in the formulas. The dimensionless evolution equations are
\begin{subequations}\label{eq:evol-non-dim}
  \begin{align}
\dot{T}=&f_{1}(T,a_{2};\chi),  & f_{1}(T,a_{2};\chi)&\equiv
\chi\left(1+\frac{3}{16}a_{2}^{\st}\right)- T^{3/2}
\left(1+\frac{3}{16}a_{2}\right), \label{eq:f1-def}\\
\dot{a_{2}}=&f_{2}(T,a_{2};\chi), &
f_{2}(T,a_{2};\chi)&\equiv
\frac{2}{T}\left[\left(T^{3/2}-\chi\right)a_{2}+B\, T^{3/2}
\left(a_2^{\st}-a_{2}\right)\right]. \label{eq:f2-def}
\end{align}
\end{subequations}

The first evolution equation governs the dynamics of the temperature: its first term, $\chi(1+3 a_{2}^{\st}/16)$ is the injected energy per unit time by the stochastic thermostat, whereas the second term, $-T^{3/2}(1+3a_{2}/16)$, is the dissipated energy per unit time by inelastic collisions. The second evolution equation governs the dynamics of the excess kurtosis and therefore the evolution of the memory variable that controls the relaxation rate at fixed temperature. Throughout this work, the aging protocol starts from the NESS corresponding to the initial temperature. Accordingly, the initial conditions for Eqs.~\eqref{eq:evol-non-dim} are
\begin{equation}\label{eq:initial-conditions}
  T_{\ini}\equiv T(t=0)=1, \quad a_{2,\ini}\equiv a_{2}(t=0)=a_2^{\st}.
\end{equation}

In the optimal-control framework developed below, the state variables $\{T,a_2\}$ define the state space of the system, while the thermostat intensity $\chi(t)$, bounded between its minimum and maximum admissible values, acts as the control parameter. The corresponding dynamics generate a set of accessible (reachable) non-equilibrium states starting from the initial NESS. The central question is therefore to identify which regions of the $(T,a_2)$ plane can be reached under admissible drivings and, among them, which states maximise or minimise the subsequent relaxation rate.


\section{Optimal preparation of the Mpemba effect}\label{sec:optimal-control}

As discussed in the introduction, the Mpemba effect arises when the cooling rate of the initially hotter sample is larger enough than that of the initially cooler, leading to the time evolution of the former overtaking that of the latter. Equation~\eqref{eq:f1-def} tells us that the cooling rate of the granular gas is given by
\begin{equation}\label{eq:cooling-rate}
   \calR_{\cool}= T^{1/2} \left(1+\frac{3}{16}a_{2}\right) ,
\end{equation}
so that the dissipated energy per unit time is $\calR_{\cool}T$. Thus, for a given value of the kinetic temperature $T$, the larger the value of the excess kurtosis $a_2$ is, the larger the cooling rate is. This is the main idea behind the pioneering analysis of the Mpemba effect in granular fluids: by tailoring the initial value of the excess kurtosis, specifically by choosing the excess kurtosis of the initially hotter sample sufficiently larger than that of the initially colder sample, the Mpemba effect emerges---see Figs.~2 and 3 of Ref.~\cite{lasanta_when_2017}. From the viewpoint developed here, the relevant question is not only which values of $a_2$ favour the Mpemba effect, but also which of those values can actually be reached under the system dynamics.

A key question, neither discussed in Ref.~\cite{lasanta_when_2017} nor in later works, is the following: what is the previous \textit{aging} protocol that leads to the non-steady states giving rise to the Mpemba effect? In the uniformly heated granular gas, the only parameter that can be externally controlled is the intensity $\chi$ of the stochastic thermostat. Therefore, one would like to elucidate how the time dependence of the thermostat intensity can be tailored in order to reach the largest and smallest accessible values of the excess kurtosis, leading to the strongest relaxation-rate asymmetry achievable between the hot and cold samples.

The discussion in the previous paragraph naturally leads to an optimal control problem. Since the thermostat intensity $\chi(t)$ is the only externally tunable parameter, the dynamics define a reachable set in the state space $(T,a_2)$ starting from the initial NESS, Eq.~\eqref{eq:initial-conditions}. The central objective is therefore to characterise the boundaries of this reachable set and establish which accessible states maximise or minimise the subsequent relaxation rate. Equivalently, we seek the optimal driving $\chi(t)$ leading to the
maximum and minimum attainable values of the excess kurtosis.


\subsection{Pontryagin's principle and transversality conditions}\label{sec:Pontryagin-and-transversality}

Let $(T,a_2)$ be the variables describing the state of the granular gas, and $\chi$ the control variable. The goal is to find a control function $\chi(t)\in[\chi_{\min},\chi_{\max}]$ that maximises or minimises the cooling rate \eqref{eq:cooling-rate}, or equivalently, the excess kurtosis $a_2$. For simplicity, we focus on the minimisation problem---the maximisation problem can be obtained from the minimisation one by changing the sign of the cost function. The optimal control problem can be stated as
\begin{equation}
    \min_{\chi(t)\in[\chi_{\min},\chi_{\max}]} a_2(t_{\rm f}),
  \label{eq:opt-control-prob-mayer}
\end{equation}
subject to the dynamics given by Eqs.~\eqref{eq:evol-non-dim} and the initial conditions \eqref{eq:initial-conditions}, where $t_{\rm f}$ is a free parameter, representing the final time, at which the cooling rate is minimised. 

It is useful to rewrite the problem in the Lagrange form, in which the cost function is a time integral instead of a terminal cost. Since the initial value of $a_2$ is fixed, the problem in Lagrange form can be equivalently written as 
\begin{equation}
  \min_{\chi(t)\in[\chi_{\min},\chi_{\max}]} \int_0^{t_{\rm f}} f_2(T,a_2;\chi)\,dt.
\label{eq:opt-control-prob-lagrange}
\end{equation}
Standard variational methods cannot be applied directly to solve the problem since the control $\chi(t)$ is bounded. Therefore, we resort to Pontryagin's Maximum Principle (PMP)~\cite{liberzon_calculus_2012}, which provides necessary conditions for optimality. 

According to PMP, if $\chi^*(t)$ is an optimal control and $(T^*(t),a_2^*(t))$ the corresponding optimal trajectory, then there exists a constant $p^*_0\leq 0$ and a function $\bm{p}^*(t)=(p_1^*(t),p_2^*(t))$, also called the costate vector, with $(p_0^*,\bm{p}^{\,*})\neq (0,0)$, such that the following conditions hold:
\begin{itemize}
    \item[a.] The optimal trajectory and the costate vector satisfy the canonical Hamilton equations
    \begin{equation}
        \dot{T}^*=\frac{\partial \mathcal{H}}{\partial p_1}, \quad
        \dot{a}_2^*=\frac{\partial \mathcal{H}}{\partial p_2}, \quad
        \dot{p}_1^*=-\frac{\partial \mathcal{H}}{\partial T}, \quad
        \dot{p}_2^*=-\frac{\partial \mathcal{H}}{\partial a_2},
        \label{eq:canonical-equations}
    \end{equation}
    where the Hamiltonian $\mathcal{H}$ is defined as
    \begin{subequations}
    \begin{align}
        \mathcal{H}(T,a_2,p_0,\bm{p},\chi)&\equiv p_1 f_1(T,a_2;\chi)+p_2 f_2(T,a_2;\chi) +p_0 f_2(T,a_2;\chi) 
        \\
        &= p_1 f_1(T,a_2;\chi)+\bar{p}_2 f_2(T,a_2;\chi),
        \label{eq:hamiltonian}
    \end{align}
    \end{subequations}
    with $\bar{p}_2 = p_0 + p_2$.
    \item[b.]The optimal control maximises the Hamiltonian as a function of the control variable for all times, i.e.
    \begin{equation}
      \mathcal{H}(T^*(t),a_2^*(t),p_0^*,\bm{p}^{\,*}(t),\chi^*(t))\geq \mathcal{H}(T^*(t),a_2^*(t),p_0^*,\bm{p}^{\,*}(t),\chi),
    \end{equation}
    for all $\chi\in[\chi_{\min},\chi_{\max}]$ and for all $t\in[0,t_{\rm f}]$.
    \item[c.] The Hamiltonian evaluated at the optimal trajectory and optimal control is zero for all times, i.e.
    \begin{equation}
      \mathcal{H}(T^*(t),a_2^*(t),p_0^*,\bm{p}^{\,*}(t),\chi^*(t))=0,
      \label{eq:hamiltonian-zero}
    \end{equation}
    for all $t\in[0,t_{\rm f}]$.
    \item[d.] The costate vector at the final time satisfies the transversality condition, i.e. the final value of the costate vector should be orthogonal to the tangent space of the set of final states. Since the final state is free, the transversality condition becomes
    \begin{equation}
      \bm{p}^*(t_{\rm f}) = 0 \Longrightarrow \{p_1^*(t_{\rm f})=0,p_2^*(t_{\rm f})=0\} \Longrightarrow \{p_1^*(t_{\rm f})=0,\ \bar{p}_2^*(t_{\rm f}) < 0\}.
      \label{eq:trans_cond}
    \end{equation}
  \end{itemize}
Note that the transversality condition, along with the condition $(p_0^*,\bm{p}^{\,*})\neq (0,0)$, prevents $p_0=0$. Hence, the strict inequality for $\bar{p}_2(t_{\rm f})$ in Eq.~\eqref{eq:trans_cond} follows. 

PMP's condition (b) entails that the arc of the optimal control in any subinterval $(t_1,t_2)\subset [0,t_{\rm f}]$ either verifies
\begin{equation}
  \left.\frac{\partial \mathcal{H}}{\partial \chi}\right|_{\chi=\chi^*(t)}=0,
\label{eq:euler-lagrange-condition}
\end{equation}
or is given by one of its bounds, i.e. $\chi^*(t)=\chi_{\min}$ or $\chi^*(t)=\chi_{\max}$. The first and second cases are usually described as a singular arc and a bang, respectively. If the optimal control does not have any singular arcs, i.e. if it alternates between the lower and upper bounds in the whole time interval $[0,t_{\rm f}]$, then it is termed a bang-bang optimal control~\cite{liberzon_calculus_2012}. 

No singular arcs are present in the solution of this optimal control problem, which thus is bang-bang---see~\ref{app:no-EL} for the rigorous proof. Since Pontryagin's Hamiltonian $\mathcal{H}$ is linear in the control function $\chi$, the bound that maximises $\mathcal{H}$ at each time instant $t$ is determined by the sign of the switching function
\begin{equation}  
\Phi(t) \equiv \frac{\partial \mathcal{H}}{\partial \chi} = p_1\left(1+\frac{3}{16}a_2^{\st}\right)-\frac{2}{T}\bar{p}_2a_2.
\label{eq:switching}
\end{equation}
Specifically, the structure of the optimal control is given by
\begin{equation}\label{eq:opt-control-structure}
  \chi^*(t) = \begin{cases}
    \chi_{\max}, & \text{if }\Phi(t) > 0, \\
    \chi_{\min}, & \text{if }\Phi(t) < 0.
  \end{cases}
\end{equation}

In our system, we have a one-bang optimal control, i.e. the optimal control $\chi(t)$ equals either $\chi_{\max}$ or $\chi_{\min}$ during the whole time window $(0,t_{\rm f})$. This stems from (i) the continuity of the switching function $\Phi(t)$, which entails that a switch between $\chi_{\min}$ and $\chi_{\max}$ may only occur at a time $t_1$ such that $\Phi(t_1)=0$, and (ii) the fact that $\Phi(t)$ only vanishes at $t=0$---see \ref{app:no-EL} for a detailed proof. Hence, the optimal control is constant in time and takes only one of the boundary values. In fact, the sign of the switching function can be related to the sign of $\bar{p}_2(0)$, which, together with Eq.~\eqref{eq:opt-control-structure}, leads to 
\begin{equation}\label{eq:optimal_control_}
  \chi^*(t) = \begin{cases}
    \chi_{\max}, & \text{if } \bar{p}_2(0)a_2^{\st} < 0, \\
    \chi_{\min}, & \text{if } \bar{p}_2(0)a_2^{\st} > 0,
  \end{cases}
\end{equation}
as proven in \ref{app:dotphi}.

Since the sign of $a_2^{\st}$ is fixed by the value of $\alpha$, choosing the sign of $\bar{p}_2(0)$  fixes the value of the optimal control for each $\alpha$. We can say more about the initial sign of $\bar{p}_2$ by analysing what happens when $\bar{p}_2=0$. Using \eqref{eq:hamiltonian-zero}, and taking into account that $p_1$ and $\bar{p}_2$ cannot vanish simultaneously (see \ref{app:no-EL} for more details), it follows that $f_1=0$. Nonetheless, the solution of the equation given the initial conditions \eqref{eq:initial-conditions} prevents this from happening---see \ref{app:zerosT}. Consequently, $\bar{p}_2\neq 0$ and thus, by the continuity of the costate vector, $\bar{p}_2$ has a fixed sign. Recalling the transversality condition \eqref{eq:trans_cond}, then $\bar{p}_2(0)<0$. Therefore, we can rewrite \eqref{eq:optimal_control_} as
\begin{equation}\label{eq:minimisation_control}
    \min_{\chi(t)\in[\chi_{\min},\chi_{\max}]} a_2 \iff \bar{p}_2(0)<0 \Rightarrow \chi^*(t) = \begin{cases}
    \chi_{\max}, & \text{if } a_2^{\st} > 0, \\
    \chi_{\min}, & \text{if } a_2^{\st} < 0.
  \end{cases}
\end{equation}


\subsection{Minimising the cooling rate}\label{sec:min-cool-rate}

We now proceed to address the minimisation of the excess kurtosis for the granular gas. Following Eq.~\eqref{eq:minimisation_control}, the sign of $a_2^{\st}$ and thus the value of the restitution coefficient $\alpha$ affects the solution of the optimal control problem. It is worth remarking that the dependence of the optimal control problem on $\alpha$ is expected from a physical point of view, since both $a_2^{\st}$ and $a_2^{\text{HCS}}$ change sign at $\alpha_c=1/\sqrt{2}$. 

Despite the Hamiltonian being linear in the control function $\chi$, the canonical equations \eqref{eq:canonical-equations} are non-linear in the state variables $\{T,a_2\}$ and thus cannot be analytically solved in closed form. As a consequence, to illustrate the optimal solution of the problem, we solve the canonical equations numerically using the optimal control in Eq.~\eqref{eq:minimisation_control}. However,  we cannot numerically solve the Hamiltonian system straightforwardly, since the initial value of the costate vector is unknown. To overcome this issue, we exploit the fact that PMP determines the costate vector only up to a multiplicative constant and thus impose a normalisation. In particular, we set $\bar{p}_2(0)=-1$ and then use Eq.~\eqref{eq:hamiltonian-zero} to obtain $p_1(0)$. With the initial costate vector sorted out, we solve the Hamiltonian system with the initial conditions given by Eq.~\eqref{eq:initial-conditions}. The integration is carried out until the transversality condition $p_1(t_{\rm f})=0$ is fulfilled, which provides the final time $t_{\rm f}$.

\begin{figure}[t]
\centering
\begin{tabular}{ccc}
\includegraphics[width=0.33\textwidth]{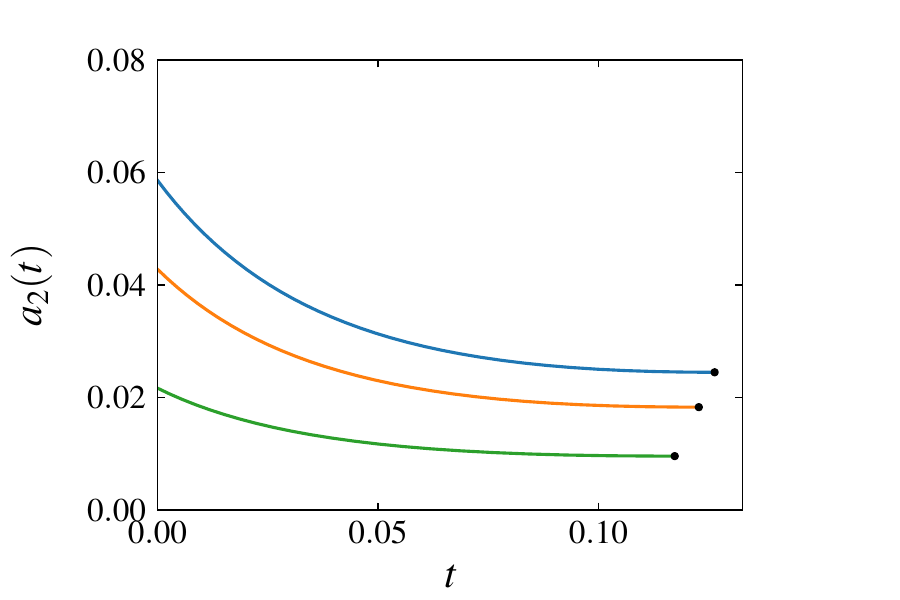} &
\includegraphics[width=0.33\textwidth]{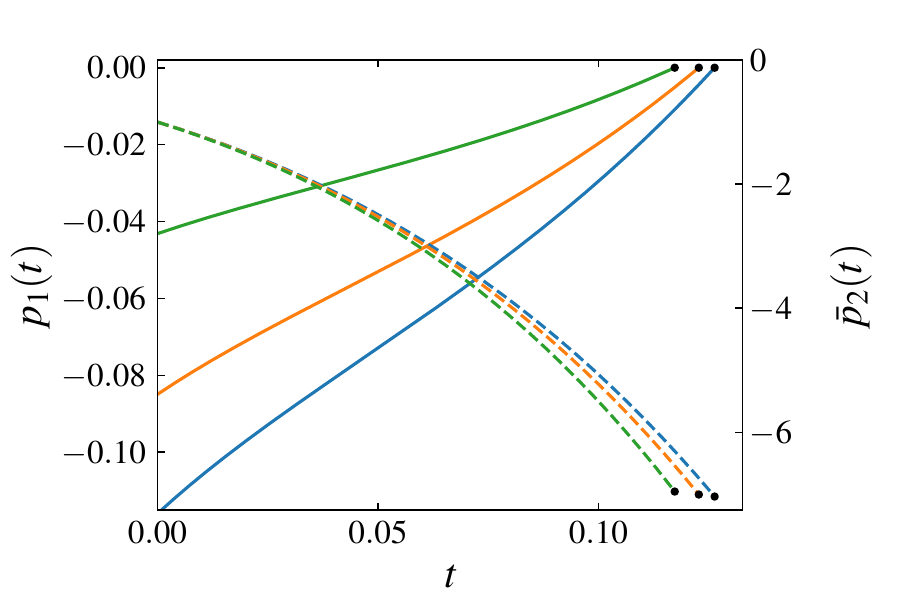} &
 \includegraphics[width=0.33\textwidth]{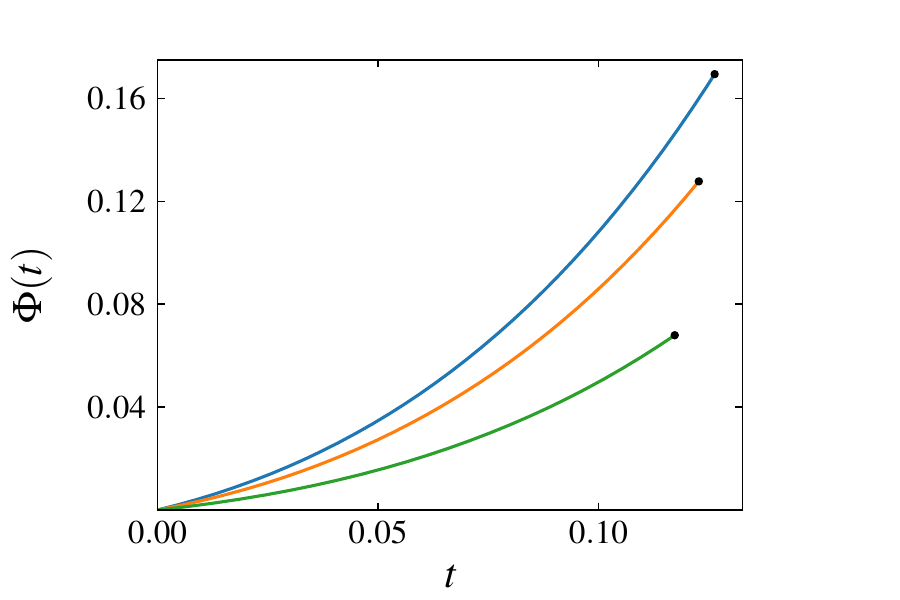}\\

\includegraphics[width=0.33\textwidth]{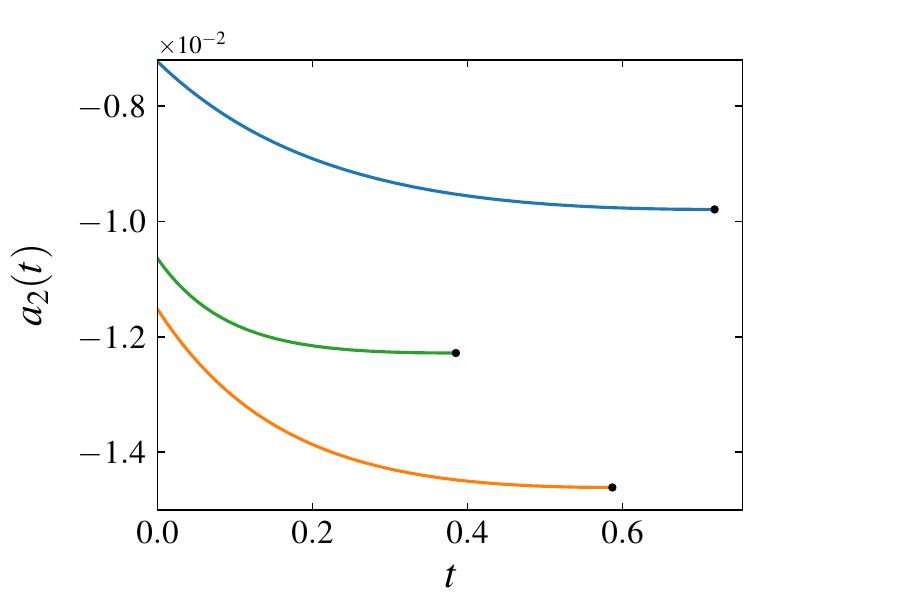} &
\includegraphics[width=0.33\textwidth]{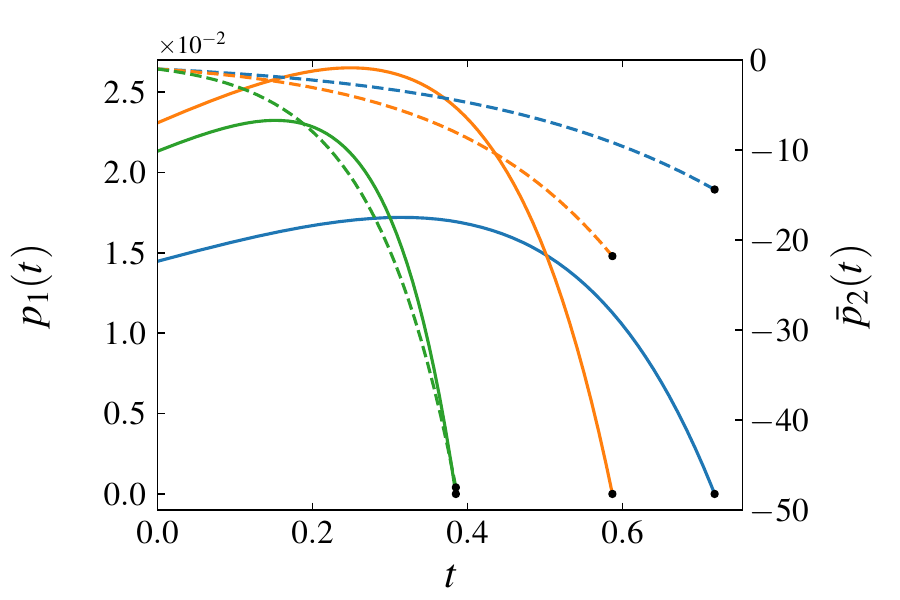} &
\includegraphics[width=0.33\textwidth]{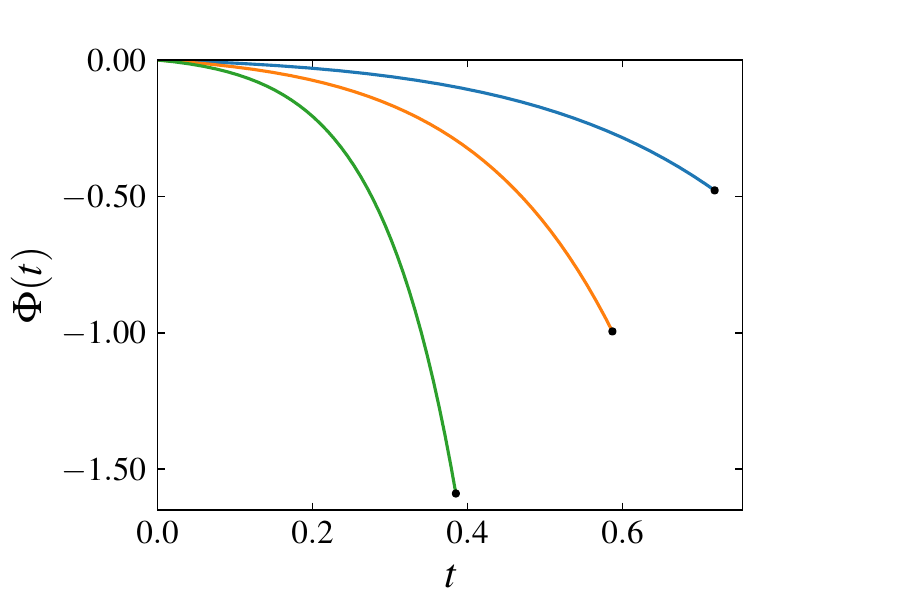}
\end{tabular}
\caption{Time evolution of the excess kurtosis $a_2$ (left), costate $\{p_1$ (solid), $\bar{p}_2$ (dashed)$\}$ (middle), and switching function $\Phi$ (right) for the optimal control with $d=3$. All curves are represented until the time $t=t_{\rm f}$ at which $a_2$ reaches its minimum. The top panels correspond to large inelasticity, $\alpha<\alpha_{\text{c}}$, with $\alpha\approx0.18$, $0.35$, and $0.53$ from top to bottom in the left panel---the same colour code is employed in the middle and right panels. Here, $\chi(t)=\chi_{\max}=10.0$ and $\bar{p}_2(0)=-1$. The bottom panels correspond to small inelasticity, $\alpha>\alpha_{\text{c}}$, with $\alpha\approx 0.78$, $0.92$ and $0.85$ from top to bottom in the bottom left panel---again, the same colour code is employed in the middle and right panels. In this case, $\chi(t)=\chi_{\min}=0.1$ and $\bar{p}_2(0)=-1$. The black points represent the value at $t=t_{\rm f}$. For both large and small inelasticity, $a_2$ evolves towards its minimum value (left panels), with the sign of $\Phi(t)$ being consistent with the value of the bang, $\chi_{\max}$ or $\chi_{\min}$, (right panels) and the costate verifying the transversality condition \eqref{eq:trans_cond} at the final time $t_{\rm f}$ (middle panels). Note that the minimum of $a_2$ is not a monotonic function of $\alpha$, which is physically reasonable---recall that neither $a_2^{\st}$ nor $a_2^{\hcs}$ are monotonic.}
\label{fig:switching-function-min}
\end{figure}

First, we consider $\alpha<\alpha_{\text{c}}$, i.e. the regime of ``large'' inelasticity---larger than that for $\alpha=\alpha_c$, in the top panels of Fig.~\ref{fig:switching-function-min}. Equation~\eqref{eq:minimisation_control} tells us that the optimal control is $\chi(t)=\chi_{\rm max}$, $\forall t\in(0,t_{\rm f})$. The consistency of the solution with PMP for different values of the (large) inelasticity is illustrated. In particular, the left panel shows the time evolution of $a_2$, which reaches its minimum at $t=t_{\rm f}$, whereas the right panel shows the switching function $\Phi(t)$, which is positive---in agreement with Eq.~\eqref{eq:opt-control-structure}. 

Second, we consider $\alpha>\alpha_{\text{c}}$, i.e. the regime of ``small'' inelasticity---smaller than that for $\alpha=\alpha_c$, in the bottom panels of Fig.~\ref{fig:switching-function-min}. Now, the optimal control is $\chi(t)=\chi_{\rm min}$, $\forall t\in(0,t_{\rm f})$. In an analogous way, the consistency of this result is illustrated: the left panel shows the evolution of $a_2$ towards its minimum value, whereas the switching function in the right panel is now negative---again, in agreement  with Eq.~\eqref{eq:opt-control-structure}.

Note that, in both cases, the numerical solution of the canonical system is consistent with our theoretical predictions for the one-bang optimal protocol: the transversality condition is fulfilled at the final time and the sign of the switching function does not change.


\subsection{Maximising the cooling rate}\label{sec:max-cool-rate}

To maximise the cooling rate, we instead minimise $-a_2(t_{\rm f})$. The problem is essentially the same as before, now with 
\begin{equation}
  \bar{p}_2(t) = p_2(t) - p_0.
\label{eq:bar-p2-max}
\end{equation}
This leads to a modification of the transversality condition, which becomes
\begin{equation}
   p_1^*(t_{\rm f})=0, \quad \bar{p}_2^*(t_{\rm f}) > 0.
  \label{eq:transversality-condition-max}
\end{equation}
Apart from this, the problem remains unchanged, so we follow the same procedure as in the previous subsection to obtain the optimal control. Since now we have $\bar{p}_2(0)>0$, the optimal control is given by
\begin{equation}\label{eq:maximisation_control}
    \max_{\chi(t)\in[\chi_{\min},\chi_{\max}]} a_2 \iff \bar{p}_2(0)>0\Rightarrow\chi^*(t) = \begin{cases}
    \chi_{\min}, & \text{if } a_2^{\st} > 0, \\
    \chi_{\max}, & \text{if } a_2^{\st} < 0.
  \end{cases}
\end{equation}

\begin{figure}[t]
\centering
\begin{tabular}{ccc}
\includegraphics[width = 0.33\textwidth]{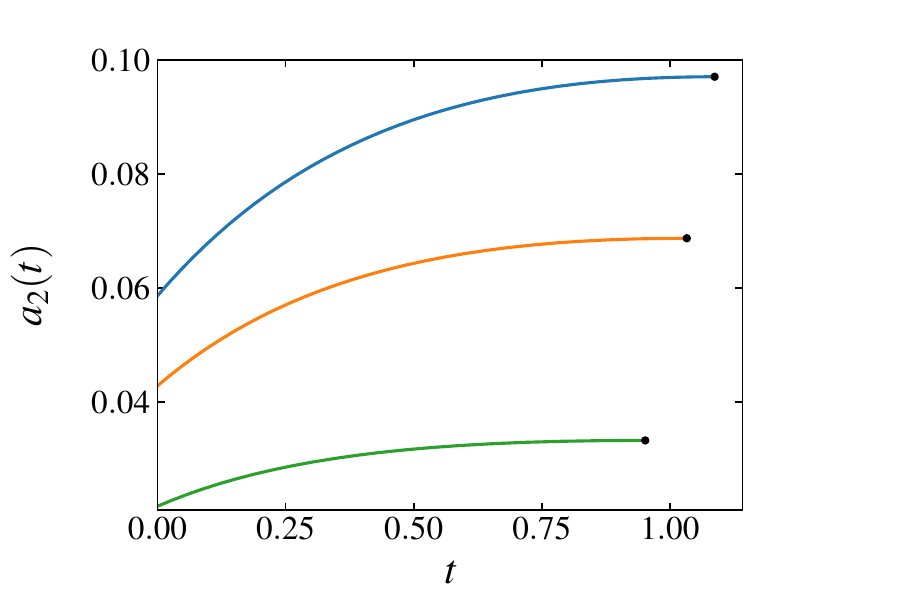} &
\includegraphics[width=0.33\textwidth]{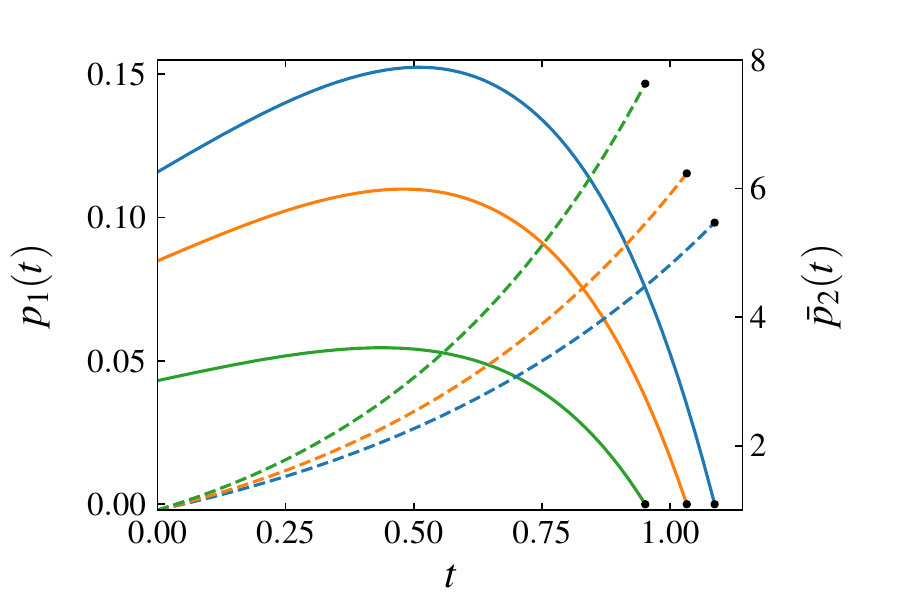} &
\includegraphics[width = 0.33\textwidth]{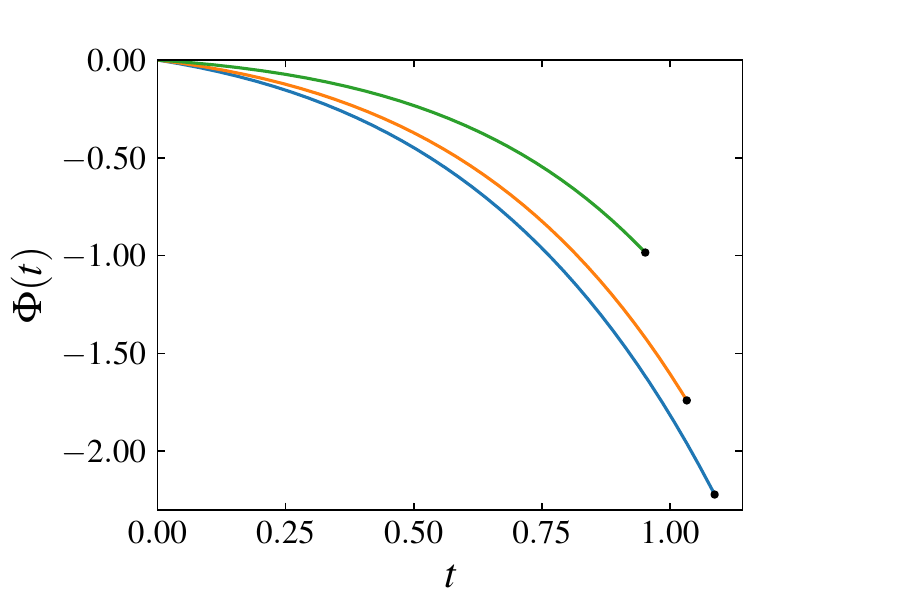}\\
\includegraphics[width = 0.33\textwidth]{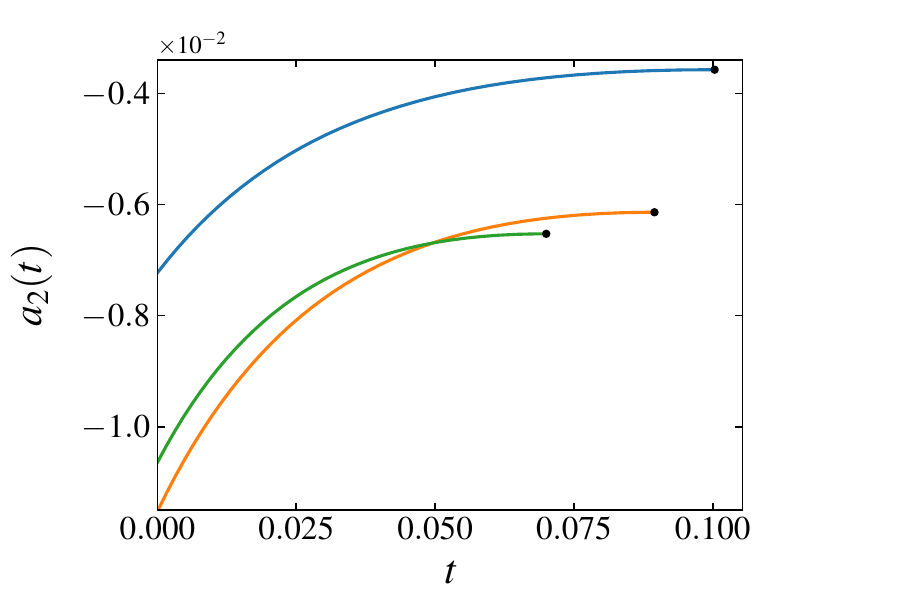} &
\includegraphics[width = 0.33\textwidth]{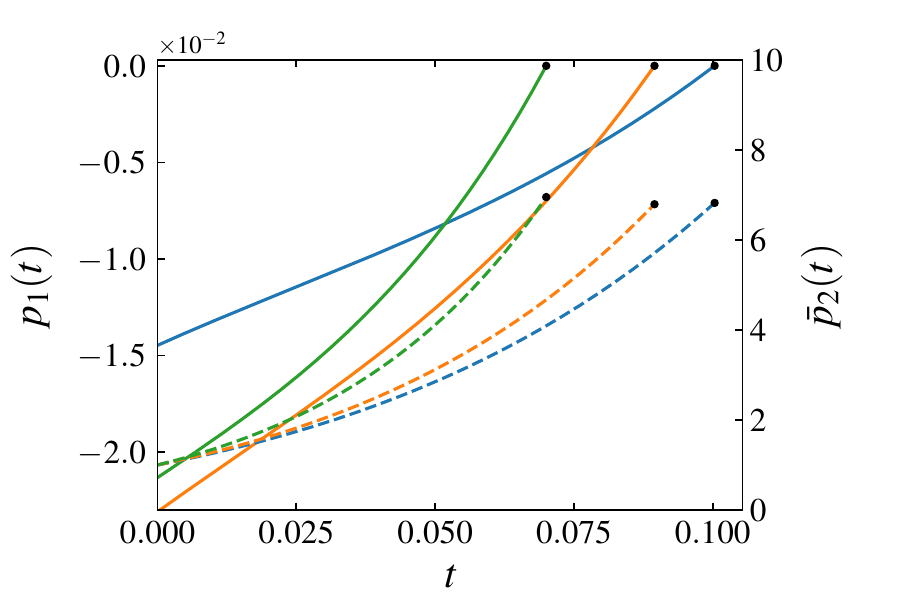} &
\includegraphics[width = 0.33\textwidth]{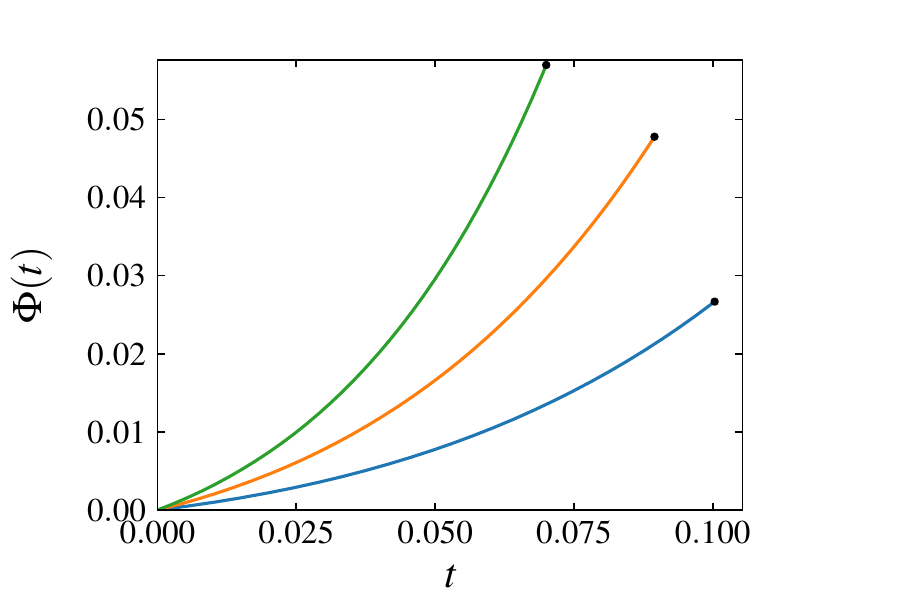} 
\end{tabular}
\caption{Time evolution of the excess kurtosis $a_2$ (left), costate $\{p_1$,$\bar{p}_2$ (dashed)$\}$ (middle), and switching function $\Phi$ (right) for the optimal control with $d=3$. All curves are represented until the time $t=t_{\rm f}$ at which $a_2$ reaches its maximum. The top panels correspond to large inelasticity, $\alpha<\alpha_{\text{c}}$, with $\alpha\approx 0.18$, $0.35$, and $0.53$ from top to bottom in the left panel---the same colour code is employed in the middle and right panels. Here, $\chi(t)=\chi_{\min}=0.1$ and $\bar{p}_2(0)=1$. The bottom panels correspond to small inelasticity, $\alpha>\alpha_{\text{c}}$, with $\alpha\approx 0.78$, $0.92$ and $0.85$ from top to bottom in the top left panel---again, the same colour code is employed in the middle and right panels. In this case, $\chi(t)=\chi_{\max}=10.0$ and $\bar{p}_2(0)=1$. The black points represent the value $t=t_{\rm f}$. For both large and small inelasticity, $a_2$ evolves towards its maximum value (left panels), with the sign of $\Phi(t)$ being consistent with the value of the bang, $\chi_{\min}$ or $\chi_{\max}$, (right panels) and the costate verifying the transversality condition \eqref{eq:transversality-condition-max} at the final time $t_{\rm f}$ (middle panels). Note that the maximum of $a_2$ is not a monotonic function of $\alpha$, which is physically reasonable---recall that neither $a_2^{\st}$ nor $a_2^{\hcs}$ is monotonic.}
\label{fig:switching-function-max}
\end{figure}

For $\alpha<\alpha_{\text{c}}$, the optimal control  is $\chi^*(t)=\chi_{\min}$, $\forall t\in(0,t_{\rm f})$. For $\alpha>\alpha_{\text{c}}$, the optimal control is $\chi^*(t)=\chi_{\max}$, $\forall t\in(0,t_{\rm f})$. Their consistency with PMP is shown in Fig.~\ref{fig:switching-function-max}.


\subsection{Ideal bounds: $\chi_{\max}\to\infty$ and $\chi_{\min}=0$}\label{sec:ideal-bounds}

The extrema for the excess kurtosis, $a_2^{\max}$ and $a_2^{\min}$, depend on the bounds for the driving $\{\chi_{\min},\chi_{\max}\}$. The looser the bounds are, i.e. the larger $\chi_{\max}$ and the smaller $\chi_{\min}$ are, the larger $a_2^{\max}$ and the smaller $a_2^{\min}$ become. It is thus especially relevant to consider the loosest possible bounds, $\chi_{\max}\to\infty$ and $\chi_{\min}=0$, which we will refer to as the ideal bounds case. Note that $\chi_{\min}=0$ is particularly simple to achieve in the granular fluid case, since it corresponds to switching off the stochastic thermostat and letting the system evolve freely. 


\subsubsection{Bang at $\chi_{\min}=0$}

In this case, the evolution equations for $\{T,a_2\}$ reduce to
\begin{equation}
  \dot{T}=- T^{3/2}
\left(1+\frac{3}{16}a_{2}\right), \quad 
\dot{a_{2}}=-2(B-1)T^{1/2} \left(a_{2}-a_2^{\hcs}\right). \label{eq:f1-f2-def}  
\end{equation}
Note that $B-1>0$ for all values of the inelasticity---see Ref.~\cite{trizac_memory_2014}. 

Since the final value of the temperature is free, $\dot{a}_2$ must vanish at the final time: this is only possible in the limit $t_{\rm f}\to\infty$, where $a_2\to a_2^\hcs$ (with $T\to 0^+$). Therefore, in the small (large) inelasticity regime, $\alpha>\alpha_c$ ($\alpha<\alpha_c$), the minimum (maximum) value of the excess kurtosis is that of the homogeneous cooling state, i.e.
\begin{align}\label{eq:a2opt-bang-zero}
    a_2^{\min}&=a_2^{\hcs} \text{ if } \alpha>\alpha_c, & a_2^{\max}&=a_2^{\hcs} \text{ if } \alpha<\alpha_c, & \chi_{\min}&=0.
\end{align}


\subsubsection{Bang at $\chi_{\max}\to\infty$}

When the upper bound $\chi_{\max}$ is very large, formally diverging, the dominant terms in the evolution equations (\ref{eq:f1-def})-(\ref{eq:f2-def}) for the bang with $\chi=\chi_{\max}$ are those proportional to $\chi$,
\begin{equation}
    \dot{T}\sim \chi_{\max} \left(1+\frac{3}{16}a_2^{\st}\right), \quad \dot{a}_2\sim -\chi_{\max}\frac{2}{T} a_2.
\end{equation}
In the bang at $\chi_{\max}\to\infty$, the system is moving along the curve~\cite{prados_optimizing_2021} 
\begin{equation}
    \frac{dT}{da_2}=-\frac{\left(1+\frac{3}{16}a_2^{\st}\right) T}{2a_2} \implies T^2 a_2^{1+\frac{3}{16}a_2^\st}={\left(a_2^\st\right)}^{1+\frac{3}{16}a_2^\st}
\end{equation}
in the $\{T,a_2\}$ plane. From the initial state $\{T_{\ini}=1,a_2^{\ini}=a_2^\st\}$, any point of this curve is reached instantaneously---in a time proportional to $\chi_{\max}^{-1}\to 0^+$. 

Now, the extremum of the excess kurtosis is $a_2=0$, corresponding to $T\to+\infty$. Therefore, in the small (large) inelasticity regime, $\alpha>\alpha_c$ ($\alpha<\alpha_c$), the maximum (minimum) value of the excess kurtosis is 
\begin{align}\label{eq:a2opt-bang-infty}
    a_2^{\max}&=0 \text{ if } \alpha>\alpha_c, & a_2^{\min}&=0 \text{ if } \alpha<\alpha_c, & \chi_{\max}&\to\infty.
\end{align}


\subsubsection{One-bang optimal protocols for the ideal case}

Putting together the above results, Eqs.~\eqref{eq:a2opt-bang-zero} and \eqref{eq:a2opt-bang-infty}, we conclude that for ideal bounds one has
\begin{align}
    \text{ if } \alpha>\alpha_c, &
    \begin{cases}
       a_2^{\min}&=a_2^{\hcs}, \\
       a_2^{\max}&=0 ;
    \end{cases}
    & 
    \text{ if } \alpha<\alpha_c, &
    \begin{cases}
       a_2^{\min}&=0, \\
       a_2^{\max}&=a_2^{\hcs};
    \end{cases}
    &  \{\chi_{\min}=0,\chi_{\max}\to\infty\}. 
\end{align}

The extrema of the excess kurtosis obtained above have a direct
geometrical interpretation in terms of the reachable set generated by
the admissible drivings. More precisely, $a_2^{\max}$ and $a_2^{\min}$
define the boundaries of the accessible non-equilibrium states in the
$(T,a_2)$ plane. Determining these extrema therefore amounts to
characterising the largest relaxation-rate asymmetry that can be
generated with the stochastic thermostat. In the ideal-bounds limit,
the accessible interval of excess kurtosis is bounded by the Gaussian
state, $a_2=0$, and the homogeneous cooling state value $a_2^{\hcs}$,
whose ordering depends on the sign of $a_2^{\hcs}$. Consequently, the stochastic thermostat imposes a fundamental reachable-state constraint on the maximum Mpemba effect that can be realised within the kinetic framework.

\begin{figure}
\centering
\includegraphics[width = 0.475\textwidth]{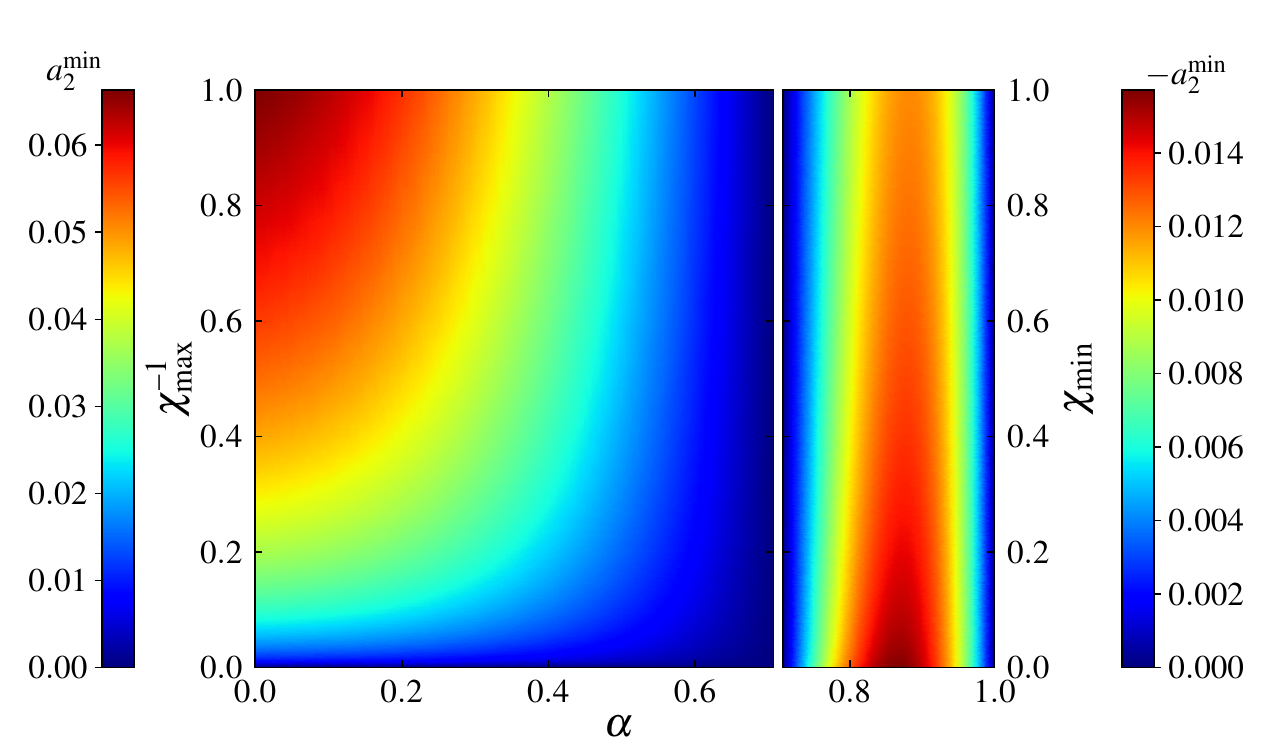}\hspace{2em}
\includegraphics[width = 0.475\textwidth]{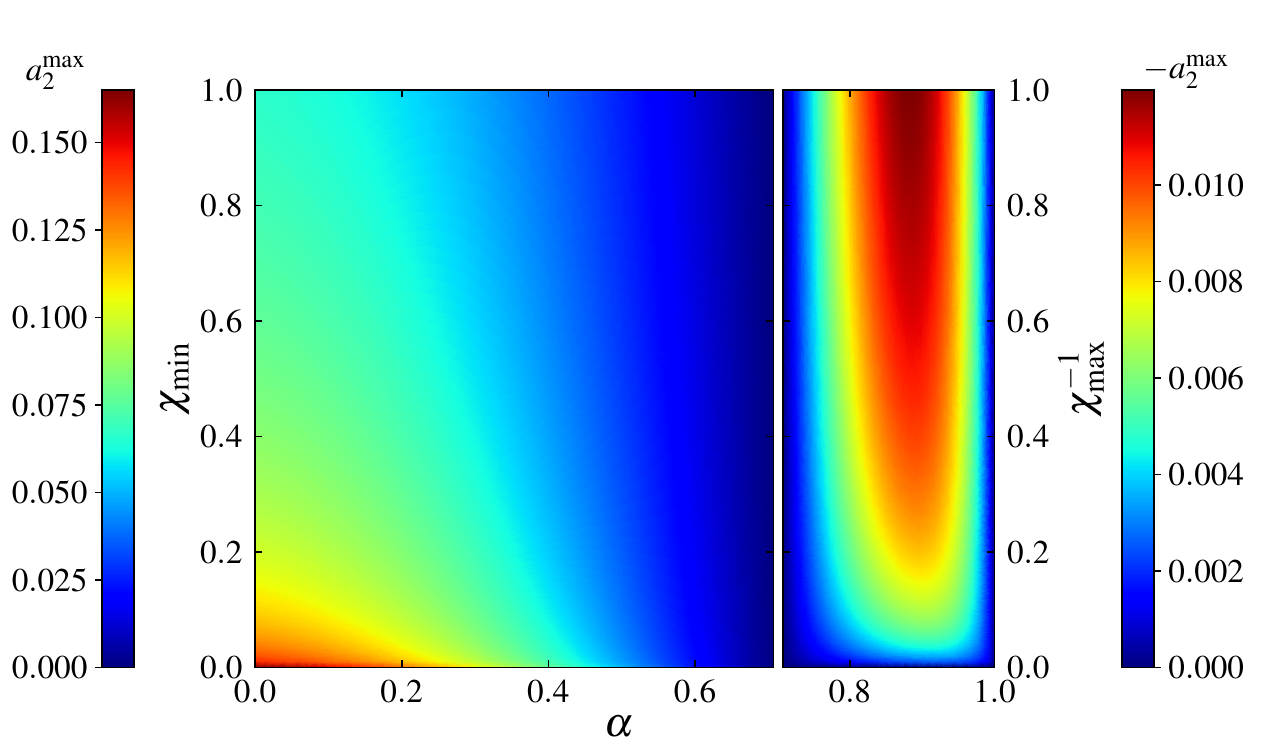}
\caption{Minimum (left panel) and maximum (right panel) accessible
values of the excess kurtosis as functions of the restitution
coefficient $\alpha$ and the control bounds
$\{\chi_{\min},\chi_{\max}\}$. These extrema define the
boundary of the reachable set generated by the stochastic
thermostat. In the left (right) panel, for $\alpha<\alpha_c$, $\chi=\chi_{\max}$ ($\chi=\chi_{\min}$), whereas for $\alpha>\alpha_c$, $\chi=\chi_{\min}$ ($\chi=\chi_{\max}$).}
\label{fig:density}
\end{figure}

Figure~\ref{fig:density} provides a map of the extremal accessible
non-Gaussianities as a function of the restitution coefficient and the
available control bounds. In this sense, it can be regarded as a
reachability diagram for the stochastic thermostat, identifying the
boundary of the accessible region in state space.

\section{Numerical simulations}\label{sec:numerics}

To test the theoretical predictions of the previous section, we have
performed numerical simulations of the granular-gas dynamics using the
Direct Simulation Monte Carlo (DSMC) method
\cite{bird_g_a_molecular_1994,brey_homogeneous_1996,montanero_monte_1996}.
Unless otherwise specified, the system consists of $N=10^6$
hard spheres ($d=3$) with unit mass, $m=1$, and unit diameter,
$\sigma=1$. The particles undergo inelastic binary collisions,
\begin{equation}
  \bm{v}_{i}^{\prime}= \bm{v}_{i}- \frac{1+\alpha}{2}
  (\bm{\hat{\sigma}}\cdot \bm{v}_{ij})
  \bm{\hat{\sigma}}, \quad
  \bm{v}_{j}^{\prime}= \bm{v}_{j}+ \frac{1+\alpha}{2}
  (\bm{\hat{\sigma}}\cdot \bm{v}_{ij})
  \bm{\hat{\sigma}},
\end{equation}
where $\bm{v}_{ij}=\bm{v}_i-\bm{v}_j$ is the relative velocity and
$\bm{\hat{\sigma}}$ is the unit collision normal. In addition, the
stochastic thermostat is implemented by applying random kicks after
every $N_c=500$ collisions. Each component of the velocity of every
particle is incremented by a Gaussian random variable with variance
$\xi^2\Delta t$, where $\Delta t$ is the time interval associated
with $N_c$ collisions
\cite{van_noije_velocity_1998,van_noije_randomly_1999,montanero_computer_2000}.

The purpose of the simulations is to test the reachable-state
constraints predicted by the optimal-control analysis. Specifically, we
verify the theoretical extrema of the excess kurtosis, which determine
the boundaries of the accessible non-equilibrium states generated by
the stochastic thermostat.

\begin{figure}[t]
\centering
\includegraphics[width = 0.49\textwidth]{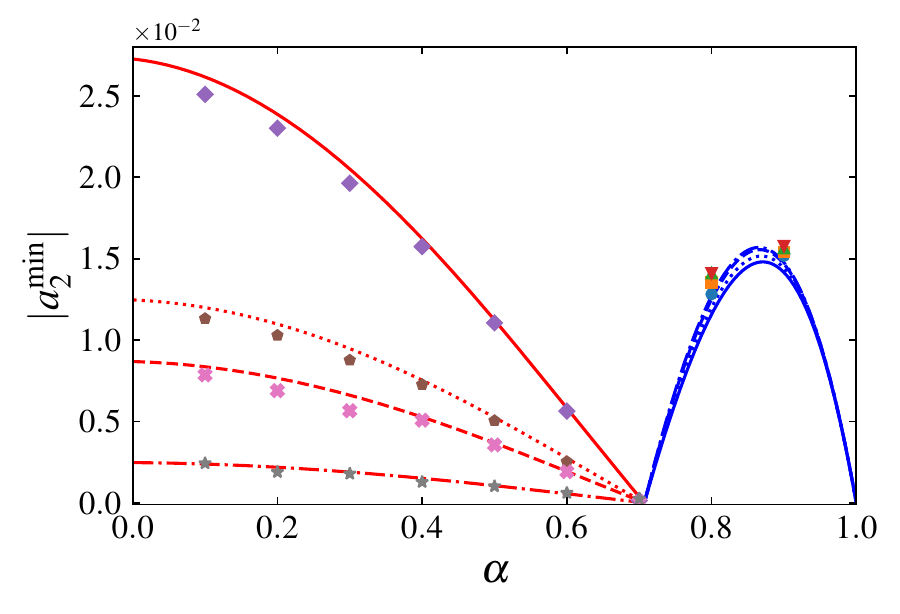}
\includegraphics[width = 0.49\textwidth]{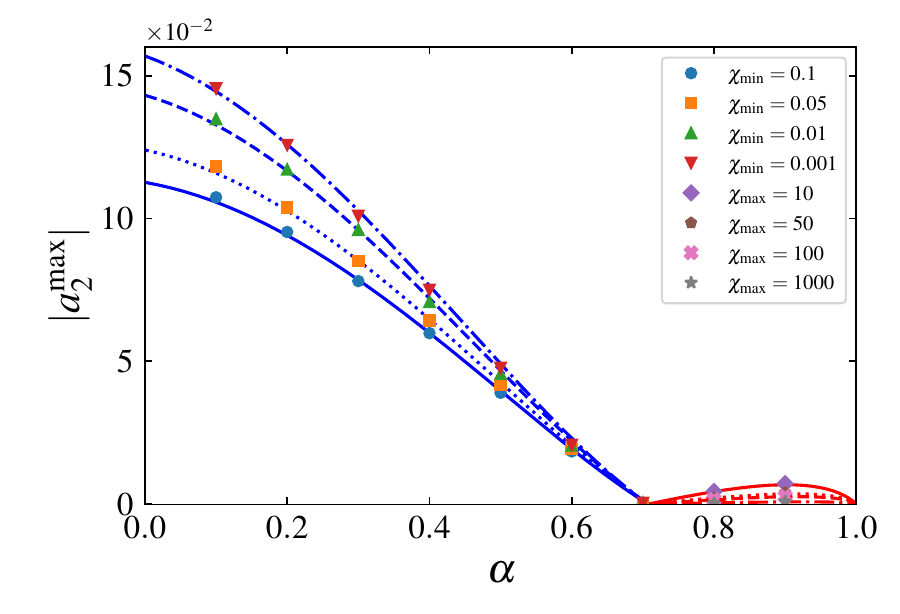}
\caption{Absolute value of the minimum (left) and maximum (right)  excess kurtosis $a_2$ as a function of the restitution coefficient $\alpha$ for different values of the control. In both panels, the red lines correspond to $\chi(t) = \chi_{\max}$, for $\chi_{\max} = 10$ (solid), $50$ (dotted), $100$ (dashed), $1000$ (dashed-dot), whereas the blue lines correspond to $\chi(t) = \chi_{\min}$, for $\chi_{\min}= 0.1$ (solid), $0.05$ (dotted), $0.01$ (dashed), $0.001$ (dashed-dot). The symbols represent the values obtained from the numerical simulation.}
\label{fig:bangvsalpha}
\end{figure}

Figure~\ref{fig:bangvsalpha} presents a global comparison between the
theoretical predictions and the DSMC results. The minimum (left panel)
and maximum (right panel) attainable values of the excess kurtosis are
plotted as functions of the restitution coefficient $\alpha$ for
different choices of the control bounds. The red curves correspond to
protocols with $\chi(t)=\chi_{\max}$, whereas the blue curves
correspond to $\chi(t)=\chi_{\min}$.

The agreement between theory and simulations is excellent throughout
the whole range of restitution coefficients and control parameters
considered. Indeed, the DSMC results validate the theoretical
predictions for the extrema $a_2^{\min}$ and $a_2^{\max}$ obtained from
the optimal-control analysis.
From the perspective developed in this work,
Fig.~\ref{fig:bangvsalpha} provides a numerical characterisation of the
reachable-state boundaries generated by the stochastic thermostat. For
finite control bounds, only a restricted interval of excess-kurtosis
values is accessible. As the bounds are progressively relaxed, i.e. as
$\chi_{\max}$ increases or $\chi_{\min}$ decreases, this interval
expands and approaches the ideal-bounds limit derived in
Sec.~\ref{sec:ideal-bounds}. The simulations clearly show the
convergence of the extrema towards the limiting values $a_2=0$ and
$a_2=a_2^{\hcs}$.

The DSMC results confirm the central prediction of the
optimal-control theory: the stochastic thermostat generates a bounded
reachable set in the $(T,a_2)$ plane whose boundaries are accurately
described by the theoretical analysis and are given, in the
ideal-bounds limit, by the Gaussian state and the homogeneous cooling
state. Consequently, the maximum relaxation-rate asymmetry---and hence
the strongest Mpemba effect attainable with this driving
mechanism---is fundamentally constrained by the accessibility of
non-equilibrium states.


\section{Discussion}\label{sec:discussion}

The central result of this work is that the strength of the kinetic
Mpemba effect cannot be discussed independently of the set of
non-equilibrium states accessible under the preparation protocol. By
formulating the aging stage as an optimal-control problem, we have
identified the extremal states reachable with a stochastic thermostat
and characterised the corresponding limits on the relaxation-rate
asymmetry. In this way, the kinetic Mpemba effect is recast as a
problem of optimal state preparation under dynamical reachability
constraints.

For uniformly heated granular gases, the accessible range of initial
conditions is bounded in a non-trivial way by the dynamics of the
thermostat. Specifically, the excess kurtosis cannot exceed, in
absolute value, the corresponding value of the homogeneous cooling
state (HCS). As a consequence, the large non-Gaussianities considered
in the seminal work of Lasanta \emph{et al}.~\cite{lasanta_when_2017}
cannot be generated using the stochastic thermostat alone. We stress
that this limitation is not fundamental, but rather reflects the
restriction imposed by the particular driving mechanism considered
here. Alternative drivings may generate a larger reachable set and thus
allow stronger relaxation anomalies.

The existence of a bounded reachable set directly constrains the
largest attainable relaxation-rate asymmetry between the hot and cold
samples. This observation is especially relevant in the context of
Pontus--Mpemba protocols, where the total time required to reach the
target state includes both the preparation and relaxation stages
\cite{nava_pontus-mpemba_2025,nava_speeding_2025,peluso_optimal_2026}.
For weak inelasticity ($\alpha>\alpha_c$), the hotter sample can at
most attain a Gaussian state ($a_2=0$), reachable through an
instantaneous strong pre-quench, whereas the colder sample is bounded
by the negative HCS value of the excess kurtosis, which is only
approached after a long cooling stage. Conversely, for strong
inelasticity ($\alpha<\alpha_c$), the hotter sample is limited by the
positive HCS value, while the colder sample can approach the Gaussian
state. Therefore, once preparation costs are incorporated into the
analysis, the relative advantage of the hotter sample is expected to be
enhanced at low inelasticity and reduced at high inelasticity.

Our results also highlight the intimate connection between the Mpemba
effect and other non-equilibrium memory phenomena. In granular fluids,
the excess kurtosis acts as an internal variable that controls both the
relaxation rate and the response to thermal history. Its importance in the present problem mirrors that in the Kovacs effect, where the amplitude and even the sign of the memory hump are also governed by the non-Gaussianities of the velocity distribution \cite{prados_kovacs-like_2014,trizac_memory_2014,patron_strong_2021}. This reinforces the view that higher-order moments of the velocity distribution encode the relevant memory variables governing the relaxation dynamics of granular gases.

Finally, several perspectives stem naturally from our work. On the one
hand, it would be of interest to extend the analysis to alternative
driving mechanisms, in order to establish how different energy-injection
protocols modify the reachable set of non-equilibrium states and the
corresponding limits on the Mpemba effect. Related questions arise in
other kinetic systems, including nonlinear fluids
\cite{santos_mpemba_2020,patron_strong_2021,patron_kinetic_2024,megias_kinetic_2022}, where the interplay between long-lived non-equilibrium states
\cite{patron_non-equilibrium_2024}, slow relaxation modes
\cite{klich_mpemba_2019,teza_eigenvalue_2023,teza_relaxation_2023},
and optimal relaxation pathways \cite{prados_optimizing_2021} remains an open problem.

On the other hand, the present work suggests a broader interpretation of
the Mpemba effect as a problem of optimal preparation under state-accessibility
constraints. This viewpoint appears naturally connected to recent
thermomajorization and resource-theoretic approaches to anomalous
relaxation, in which the occurrence of the Mpemba effect is linked to
admissible state transformations
\cite{van_vu_thermomajorization_2025,summer_resource-theoretical_2026}.
It may also prove useful in quantum systems, where accelerated
relaxation has been related to the spectral structure of the dynamics
\cite{klich_mpemba_2019,ares_quantum_2025}. More generally, our results
suggest that anomalous relaxation, optimal control, and state
accessibility are deeply interconnected, and that engineering
relaxation amounts to navigating the geometry of the reachable
non-equilibrium state space.


\section*{Acknowlegments}
We acknowledge financial support from Grant PID2024-155268NB-I00 funded by MICIU/AEI/10.13039/5011000\hspace{0pt}11033/ FEDER, UE. We also acknowledge support from the applied research and innovation Project PPIT2024-31833, cofunded by EU--Ministerio de Hacienda y Función Pública--Fondos Europeos--Junta de Andalucía--Consejería de Universidad, Investigación e Innovación. A.P.C. also acknowledges support from the Natural Sciences and Engineering Research Council of Canada (NSERC) Discovery Grant and Discovery Accelerator Supplement RGPIN-2020-04950. C.R.M. acknowledges support from the Ministry of Science, Innovation and Universities FPU program through Grant No. FPU22/01151.


\appendix


\section{Zeros of the switching function}
\label{app:no-EL}

Here, we study the conditions under which the switching function is zero. We prove that, along an optimal trajectory of the problem \eqref{eq:opt-control-prob-mayer}, the switching function cannot be zero at any time $t\neq0$. Consequently, there are no singular arcs in the optimal control, and it only consists of a single bang arc.

Let us assume that for $t_1\in[0,t_{\rm f}]$, $\Phi(t_1)=0$. From the definition of the switching function \eqref{eq:switching}, we can express $p_1$ in terms of $\bar{p}_2$ at $t=t_1$ as
\begin{equation}
    p_1 = \frac{2a_2}{T\left(1+\frac{3}{16}a_2^{\st}\right)}\bar{p}_2.
    \label{eq:appen_p1p2}
\end{equation}
Substituting \eqref{eq:appen_p1p2} into condition \eqref{eq:hamiltonian-zero} we obtain
\begin{equation}
    \mathcal{H}=\frac{2\bar{p}_2T^{1/2}(a_2^{\st}-a_2)}{1+\frac{3}{16}a_2^{\st}}\left[\frac{3}{16}a_2+B\left(1+\frac{3}{16}a_2^{\st}\right)\right]=0.
    \label{eq:append_hzero}
\end{equation}
Since $T>0$, if \eqref{eq:append_hzero} holds, we have three possibilities:
\begin{enumerate}
    \item $\bar{p}_2=0$:
    
    In this case, \eqref{eq:appen_p1p2} yields $p_1=0$. Since the costate dynamics are governed by a homogeneous linear differential system, this condition means that $p_1=0, \bar{p}_2=0$ for all times. However, that is incompatible with the transversality condition \eqref{eq:trans_cond}, where $\bar{p}_2(t_f)\neq0$.   
    \item $a_2 = a_2^{\rm B} \equiv -\frac{16}{3}\left(1+\frac{3}{16}a_2^{\st}\right)B$:
    
    This value of $a_2$, however, is not possible. To prove this, we first derive a lower bound of the excess kurtosis $a_2$ using the non negativity of the variance of the velocity, and the definition of $a_2$ \eqref{eq:T-a2-def},
    \begin{equation}
        \sigma^2_{v^2}=\langle v^4\rangle - \langle v^2\rangle^2\geq0 \Rightarrow a_2\geq -\frac{2}{d+2},
        \label{eq:a2lowerbound}
    \end{equation}
    Applying \eqref{eq:a2lowerbound} and $B>1$, we obtain
    \begin{equation}
        a_2^{\rm B} \leq -\left(\frac{16}{3}-\frac{2}{d+2}\right)B\leq -\left(\frac{16}{3}-\frac{2}{d+2}\right)\leq -\frac{2}{d+2}\left(\frac{16d+26}{3}\right)<-\frac{2}{d+2}.
    \end{equation}
    However, this contradicts \eqref{eq:a2lowerbound}.
    \item $a_2=a_2^{\st}$: 

    To discard this case for all $t\neq 0$, let us first assume that there exists an interval $(t_1,t_2)\subseteq[0,t_f]$ such that $a_2(t)=a_2^{\st}$ for all $t\in(t_1,t_2)$. This implies that $a_2=a_2^{\st}$, and therefore $f_2(t)=0$, which yields $T=\chi^{2/3}$ for all $ t\in(t_1,t_2)$. Hence, using \eqref{eq:f1-def}, we conclude that $T$ is also constant in $(t_1,t_2)$. This trajectory keeps the system unchanged, and therefore, cannot be part of an optimal solution. 

    The only remaining possibility is that there exists some $t'\in(0,t_f]$ such that $a_2(t)=a_2^{\st}$. Let us assume that $t'$ is the earlier instant at which this occurs. Then, for all $t\in[0,t']$, the control is either $\chi_{\min}$ or $\chi_{\max}$, and is therefore constant. Since the control is constant, and hence continuous, it follows from \eqref{eq:f2-def} that $f_2$ is continuous as well. Moreover, because $t_f$ is defined by the condition $f_2=0$ through the transversality condition \eqref{eq:trans_cond}, it must have a definite sign on $[0,t')$. Since $a_2(0)=a_2^{\st}$, this implies $t'>t_f$ and therefore, $a_2(t)\neq a_2^{\st}$ for all $t\in(0,t_f]$.   
\end{enumerate}
Putting all together, we can conclude that,
\begin{equation}
    \Phi(t)\neq 0 \quad \forall t\in(0,t_{\rm f}].
\end{equation}
Therefore, singular arcs are strictly excluded from the optimal solution. Since the switching function is continuous and has no zeros for $t>0$, its sign cannot change along an optimal trajectory. Consequently, every optimal solution consists of a single bang arc.


\section{Sign of the switching function}\label{app:dotphi}

Here, we analyse the sign of the switching function through the study of the initial sign of its time derivative. Since the switching function vanishes only at the initial time, this sign fixes the sign of the switching function along the optimal trajectory and therefore characterises which value of the bang corresponds to the optimal control.

The time-derivative of the switching function is
\begin{equation}
    \dot{\Phi} = \left(1+\frac{3}{16}a_2^{\st}\right)\dot{p}_1 - \frac{2a_2}{T}\dot{\bar{p}}_2 - \frac{2\bar{p}_2}{T}\dot{a}_2 + \frac{2\bar{p}_2a_2}{T^2}\dot{T}.
\end{equation}
Using Eqs.~\eqref{eq:canonical-equations} and \eqref{eq:hamiltonian-zero} and evaluating them at $t=0$, where $T(0)=1$ and $a_2(0)=a_2^{\st}$, we obtain
\begin{equation}
    \dot{\Phi}(0) = -\frac{4\bar{p}_2(0)a_2^{\st}}{1+\frac{3}{16}a_2^{\st}}\left[\frac{3}{16}a_2^{\st}+B\left(1+\frac{3}{16}a_2^{\st}\right)\right].
    \label{eq:dotphi}
\end{equation}
Using the lower bound of $a_2$ \eqref{eq:a2lowerbound} and $d\geq1$,
\begin{equation}
    a_2\geq-\frac{2}{d+2}\geq -\frac{2}{3}
\end{equation}
Hence,
\begin{equation}
    \frac{3}{16}a_2^{\st}+B\left(1+\frac{3}{16}a_2^{\st}\right)\geq -\frac{1}{8} + \frac{7}{8}B = \frac{7B-1}{8}>\frac{3}{4}>0.
    \label{eq:parenth}
\end{equation}
where we have used that $B>1$. Therefore, all factors in \eqref{eq:dotphi} except $p_2(0)$ and $a_2^{\st}$ are always positive. Consequently, the sign of $\dot{\Phi}(0)$ is given by
\begin{equation}
    \operatorname{sgn}\{\dot{\Phi}(0)\} = -\operatorname{sgn}\{\bar{p}_2(0)a_2^{\st}\}.
\end{equation}
Following \ref{app:no-EL}, the switching function vanishes only at the initial time, \(t=0\). Since \(\Phi\) is continuous and \(\dot{\Phi}(0)\neq0\), the sign of \(\Phi(t)\) for \(t>0\) is fixed by \(\dot{\Phi}(0)\). Thus, the sign of $\dot{\Phi}(0)$ uniquely determines the sign of
$\Phi(t)$ for all $t>0$ and therefore fixes which of the two
admissible bangs constitutes the optimal control, as given by Eq.~\eqref{eq:optimal_control_}.


\section{$\dot{T}\neq 0$ along an optimal trajectory}
\label{app:zerosT}

The purpose of this appendix is to establish that the temperature
cannot remain stationary along an optimal trajectory. This property is
used in Sec.~\ref{sec:optimal-control} to identify the sign of the
costate variable $\bar p_2$.

We prove that, under the initial conditions
\eqref{eq:initial-conditions}, one has $\dot{T}\neq0$ for all
$t\in(0,t_{\rm f})$ along an optimal trajectory. Recall that the
optimal control is constant and equal to one of its bounds throughout
the interval $(0,t_{\rm f})$. Since $t_{\rm f}$ is determined by the
transversality condition---Eq.~\eqref{eq:trans_cond} for the
minimisation problem and Eq.~\eqref{eq:transversality-condition-max}
for the maximisation problem---the continuity of $\chi$ and therefore
of $\dot a_2$ implies that $\dot a_2$ has a definite sign along the
optimal trajectory.

Using the initial condition $a_2(0)=a_2^{\st}$, its sign can be
written as
\begin{equation}
    \operatorname{sgn}\{\dot{a}_2(t)\}
    =
    -\operatorname{sgn}\{a_2^{\st}-a_2(t)\}.
    \label{eq:signa2}
\end{equation}
Assume now, by contradiction, that there exists a time
$\hat t\in(0,t_{\rm f})$ such that $\dot T(\hat t)=0$. Using
Eq.~\eqref{eq:f1-def}, one obtains
\begin{equation}
    \hat{\chi}\left(1+\frac{3}{16}a_{2}^{\st}\right)
    -
    \hat{T}^{3/2}\left(1+\frac{3}{16}\hat a_2\right)
    =0,
\end{equation}
which implies
\begin{equation}
    \hat{\chi}
    =
    \hat{T}^{3/2}
    \frac{1+\frac{3}{16}\hat a_2}
         {1+\frac{3}{16}a_2^{\st}}.
\end{equation}
Substituting this expression into Eq.~\eqref{eq:f2-def} yields
\begin{equation}
    \dot a_2(\hat t)
    =
    \frac{2\hat T^{1/2}}
         {1+\frac{3}{16}a_2^{\st}}
    \left(a_2^{\st}-\hat a_2\right)
    \left[
        \frac{3}{16}\hat a_2
        +
        B\left(1+\frac{3}{16}a_2^{\st}\right)
    \right].
\end{equation}

Using the lower bound \eqref{eq:a2lowerbound} together with $B>1$, it
follows that
\begin{equation}
    1+\frac{3}{16}a_2^{\st}>0,
\end{equation}
and
\begin{equation}
     \frac{3}{16}\hat a_2
     +
     B\left(1+\frac{3}{16}a_2^{\st}\right)
     >
     0.
\end{equation}
Therefore,
\begin{equation}
    \operatorname{sgn}\{\dot a_2(\hat t)\}
    =
    \operatorname{sgn}\{a_2^{\st}-\hat a_2\}.
\end{equation}
This contradicts Eq.~\eqref{eq:signa2}. 

Hence, there cannot exist any
$\hat t\in(0,t_{\rm f})$ such that $\dot T(\hat t)=0$, i.e.
\begin{equation}
    \dot T(t)\neq0,
    \qquad
    \forall\,t\in(0,t_{\rm f}),
\end{equation}
which proves the claim.


\bibliographystyle{elsarticle-num} 
\bibliography{Mi-biblioteca-26-may-2026}

\end{document}